\documentclass[preprint,12pt]{elsarticle}
\usepackage[utf8]{inputenc}
\usepackage[T1]{fontenc}

\usepackage{amsmath}
\usepackage{amsfonts}
\usepackage{amssymb}
\usepackage{amsthm}
\usepackage{color}
\usepackage{url}
\usepackage{subfig}
\usepackage{verbatim}

\usepackage{graphicx} 
\usepackage{realboxes}

\renewcommand{\vec}[1]{{\mathbf{#1}}}
\newcommand{\eps}{\varepsilon}
\newcommand{\Eins}{\mathbf{1}}
\newcommand{\dx}{\,\mathrm{d}x}
\newcommand{\dA}{\,\mathrm{d}(x,y)}
\title{A conservative discontinuous Galerkin scheme \\ for the 2D incompressible Navier--Stokes equations}
\journal{Computer Physics Communications}
\author[maths]{L.~Einkemmer}
\ead{Lukas.Einkemmer@uibk.ac.at}
\author[physics]{M.~Wiesenberger}
\ead{Matthias.Wiesenberger@uibk.ac.at}
\address[maths]{Department of Mathematics, University of Innsbruck, Austria}
\address[physics]{Institute for Ion Physics and Applied Physics, Association
  Euratom-\"OAW, University of Innsbruck, Austria} 

\begin{document}

\begin{abstract}
    In this paper we consider a conservative discretization of the
    two-dimensional incompressible Navier--Stokes equations. 
    We propose an extension of Arakawa's classical finite difference scheme
    for fluid flow in the vorticity-stream function formulation to a high
    order discontinuous Galerkin approximation.
In addition, we show numerical simulations that demonstrate the accuracy of
    the scheme and verify the conservation properties, which are essential
    for long time integration.
Furthermore, we discuss the massively parallel implementation on graphic processing units.
\end{abstract}

\begin{keyword}
    Arakawa's method \sep discontinuous Galerkin \sep incompressible Navier--Stokes equations \sep conservative methods, two-dimensional fluids
\end{keyword}

\maketitle

\section{Introduction}

The purpose of this paper is to develop a novel conservative space 
discretization for the two-dimensional incompressible Navier--Stokes equations
\begin{subequations}
\begin{align}
	\partial_t \omega + \{\psi, \omega\} = D\Delta \omega  \\
	-\Delta \psi = \omega,
\end{align}
\label{eq:vorticity}
\end{subequations}
where the Poisson bracket is given by $\{\psi, \omega\}=\psi_x \omega_y - \psi_y \omega_x$, $\omega$ is 
the vorticity and $\psi$ the streamfunction.

In 1966 Arakawa introduced a finite difference approximation that conserves the single linear (vorticity) and the two quadratic invariants (kinetic energy and enstrophy) of equation \eqref{eq:vorticity} for $D=0$. The conservation properties of this scheme are important as they enable the long time integration of this equation without introducing numerical instabilities and other unphysical artifacts. 

Despite these advantages, Arakawa's scheme was not appreciated in many
physical applications because it is not generalizable to three
dimension. However, in recent years it has received increasing attention 
from the plasma physics community. 
In magnetized plasmas the time- and length-scales of low-frequency drift wave
dynamics parallel and perpendicular to the magnetic field can be separated. 
The fluid-like advection and transport by drift wave turbulence is thus,
essentially, a quasi two-dimensional phenomenon. 
This was exploited early, e.g.~by the Hasegawa-Wakatani standard model 
\cite{Hasegawa1983} for dissipative drift wave turbulence. 
Also in more recent gyrofluid models \cite{Scott2010} the advection 
operators split into derivatives along and perpendicular to the magnetic
field line, making two-dimensional Poisson brackets reappear in a 
three dimensional model. The numerical simulation of these equations
is a challenging task and there is a need for efficient methods
that are parallelizable and accurate.  

In this context Arakawa's scheme has been re-considered in
ref.~\cite{Naulin2003} and is regarded as preferable compared to specific spectral and
finite difference schemes with respect to conservation properties.
In ref.~\cite{Peterson2013} the Arakawa method has been compared to
higher-order extrema preserving upwind methods, which in contrast are designed
to ensure positivity. Positivity preserving methods would be favorable for
applications where strong variations in advected quantities appear, like in
shocks or in scrape-off layer fusion plasmas, whereas energy and enstrophy
preservation is essential to ensure accuracy of long time scale phenomena,
like the generation of zonal flows and mean equilibrium shear flows in the
edge of fusion plasmas. 

In recent years discontinuous Galerkin (dG) methods have been investigated as an excellent alternative to finite difference and finite volume schemes in numerical simulations involving both parabolic as well as hyperbolic problems (for the advection dominated case see, for example, the review article \cite{cockburn2001runge}). Such methods combine many advantages of finite element methods (such as the ease of handling complicated geometry) with properties more commonly associated with finite difference approximations. Examples of the latter includes the absence of a global mass matrix. In addition, the nearest neighbor character of high order dG discretizations facilitates an efficient parallel implementation. 

Our goal in this paper is therefore to extend the scheme proposed by Arakawa to a high order dG method. That is, we propose a dG method that conserves the vorticity, kinetic energy, and enstrophy and show that such a scheme exhibits significant gains in accuracy as compared to finite difference approximations. In addition, we demonstrate the feasibility of efficient parallelization by providing an implementation on graphic processing units (GPUs).

The outline of this paper is as follows. 
In section \ref{sec:arakawa} we provide a derivation of the finite difference scheme introduced in \cite{arakawa1966} without explicitly using the conditions imposed by the conservation of the linear and quadratic invariants. 
This then allows us to extend the before mentioned approach to dG methods, 
resulting in a conservative dG scheme (section \ref{sec:dG}). 
In section \ref{sec:implementation} we derive the coefficients of the discrete derivative matrices 
by employing orthogonal Legendre-polynomials as the basis. The implementation
thereof is briefly described in section
\ref{sec:simulation},
where we also emphasize the high parallel efficiency of our dG methods by showing results from CPU and GPU run time measurements. 
We then numerically verify the previously claimed conservative 
properties of our newly derived scheme. 
In sections \ref{sec:simulation} and \ref{sec:simulation2} we present results from 
the time-integration of the two-dimensional incompressible Navier--Stokes
equations. We evaluate the order of our discretization and 
examine the conservation properties using the
well known Lamb dipole solution \cite{Nielsen1997}. 
Finally, we conclude in section~\ref{sec:conclusion}.

\section{Arakawa's scheme as a consequence of the product rule} \label{sec:arakawa}
In the absence of viscosity,  
equation (\ref{eq:vorticity}) 
conserves the total vorticity $V$, the kinetic energy $E$, and 
the enstrophy $\Omega$, as given by
\begin{align}
    V:=\int_C \omega\dA,\ E:=\frac{1}{2}\int_C (\nabla\psi)^2\dA,\ \Omega :=\frac{1}{2}\int_C \omega^2\dA,
    \label{eq:quantities}
\end{align}
if periodic or impermeable walls are imposed on the boundary of the domain $C$.
Note that Green's formula yields the following practical identity for the energy \mbox{$E=\tfrac{1}{2}\int_C \psi\omega\dA$}.
The conservation of vorticity is derived by integrating 
over the whole domain. 
\begin{align}
    \frac{\partial}{\partial t}V &=  \int_{C} \dot\omega \dA  = \int_C \{\omega, \psi\}\dA = 0.
    \label{}
\end{align}
For the last equality we integrate by parts and use the boundary conditions.
We derive the conservation of $E$ and $\Omega$ by 
multiplying equation (\ref{eq:vorticity}) by $\psi$ and $\omega$ respectively
(before conducting the integration). This yields
\begin{subequations}
\begin{align}
    \frac{\partial}{\partial t}E &=  \int_{C}\psi\dot\omega \dA = \int_C \psi \{\omega, \psi\} \dA = 0 \\
    \frac{\partial}{\partial t}\Omega &= \int_{C} \omega\dot\omega \dA = \int_C \omega \{\omega, \psi\} \dA = 0.
\end{align}
\end{subequations}
Here we have used the product rule on the right hand side
and integration by parts to show the last identity.

In the seminal paper by Arakawa \cite{arakawa1966}  a finite difference approximation of the Poisson bracket $\{f,g\}=f_x g_y - f_y g_x$ is constructed such that the discrete versions of the integrals
\begin{subequations}
\begin{align}
	\int_{C} \{f,g\} \dA&=0\label{eq:average1} \\
	\int_{C} f\{f,g\}\dA&=0\label{eq:average2} \\
	\int_{C} g\{f,g\}\dA&=0\label{eq:average3}
\end{align}
\end{subequations}
hold true.  
These requirements imply, that $V$ is exactly conserved
in a linear time integration scheme,
while $E$ and $\Omega$ are conserved up to errors of the time-stepping scheme only.

Arakawa proceeds by determining conditions on the coefficients of a general finite difference scheme such that the constraints given by equations \eqref{eq:average1}-\eqref{eq:average3} are satisfied.
However, we will introduce a different approach here. This approach provides a simpler construction of the finite difference scheme obtained in \cite{arakawa1966} and makes the generalization to the dG methods considered in this paper straightforward. 

To that end let us remark that in the continuous case, by using the product rule and integration by parts as shown above, the constraints \eqref{eq:average1}-\eqref{eq:average3} are immediately satisfied, if periodic or homogeneous Dirichlet boundary conditions are assumed. However, contrary to integration by parts, the product rule is no longer true, in general, if the constraints are discretized. 
This is most easily demonstrated with second order centered differences. In this case, we have
\begin{equation*}
	(fg)_x \approx \frac{f_{i+1}g_{i+1}-f_{i-1}g_{i-1}}{2h} \neq 
	\frac{f_{i+1}g_i-f_{i-1}g_i+f_ig_{i+1}-f_ig_{i-1}}{2h} \approx f_xg + fg_x,
\end{equation*}
where we have used $h$ to denote the cell size.

We now argue that the violation of the product rule lies at the heart of the problem, as instead of $\{f,g\}=f_xg_y-f_yg_x$ we could just as well discretize $\{f,g\}=(fg_y)_x-(fg_x)_y$ or $\{f,g\}=(f_xg)_y-(f_yg)_x$. All of these representations are equivalent in the continuous case, by virtue of the product rule, but are, in general, different if a space discretization is considered. Thus, instead of choosing a single representation let us use an equal superposition, i.e.
\begin{subequations}
\begin{align}
	J &:= \tfrac{1}{3}(J^{++} + J^{+x} + J^{x+}) \\
	J^{++} &:= D_x(f) D_y(g) - D_y(f) D_x(g) \\
	J^{+x} &:= D_x(f D_y(g)) - D_y(f D_x(g)) \\
	J^{x+} &:= D_y(D_x(f) g) - D_x(D_y(f) g),
\end{align}
\label{eq:J}
\end{subequations}
where $D_x$ and $D_y$ denote the finite difference operator in the $x$ and $y$-direction, respectively. It is straightforward to see why this construction works. For the purpose of this demonstration we consider equation \eqref{eq:average2}, which once discretized by the above scheme can be written as
\begin{equation*}
	\int_C f (J^{++}+J^{+x}+J^{x+}) \,\mathrm{d}(x,y) = 0,
\end{equation*}
where the integral in the finite difference case is defined as the appropriate summation weighted by the cell size. 
Using integration by parts (but not the product rule) and neglecting boundary terms, we find at once that 
\begin{equation*}
	\int_C f (J^{++}+J^{+x}) \,\mathrm{d}(x,y) = 0.
\end{equation*}
Thus, we are left with
\begin{align*}
	&\int_C f D_y(D_x(f)g) \,\mathrm{d}(x,y) - \int_C f D_x(D_y(f)g) \,\mathrm{d}(x,y) \\
	&\qquad = - \int_C D_y(f) D_x(f)g \,\mathrm{d}(x,y) + \int_C D_x(f) D_y(f)g \,\mathrm{d}(x,y) = 0,
\end{align*}
which yields zero, as is necessary to satisfy equation \eqref{eq:average2}.

Note that the scheme given by equation \eqref{eq:J} is exactly the scheme obtained by Arakawa. However, it has been constructed without explicitly deriving rather tedious conditions on the coefficients of the finite difference scheme. 
In \cite{arakawa1966} an additional term, denoted by $J_{ij}^{xx}$, is considered. It results from the space discretization of  
\begin{equation*}
	(f_x+f_y)(g_y-g_x)-(f_y-f_x)(g_x+g_y),
\end{equation*}
but it is found that the coefficient of this term vanishes, if conservation is to be achieved. This, however, is evident in our framework as it is not connected to the product rule (even though it clearly does satisfy the order conditions and constitutes a valid approximation of second order).

\section{A conservative discontinuous Galerkin approximation} \label{sec:dG}
In the previous section we have demonstrated that the failure of the classic finite difference approximations to conserve certain invariants can be analyzed by considering the violation of the product rule in the discrete setting. This approach then leads to an alternative way of deriving the conservative finite difference scheme discovered by Arakawa.

Let us now employ a dG method to discretize the problem in space. To that end let us approximate a function $f(x)$ by projecting it on the space generated by a set of orthonormal basis functions $p_{n0}, p_{n1}, \dots, p_{n,P-1}$, where the cell index is denoted by $n$. The coefficients $f^{ni}$ are then given by
\begin{equation*}
	f^{ni} = \int_{C_n} f(x)p_{ni}(x) \,\mathrm{d}x,
\end{equation*}
where the $n$-th cell is defined as $C_n = [x_{n-1/2},x_{n+1/2}]$. The extension to two dimensions is immediate for a tensor product grid; therefore, we will restrict the discussion in this section to the case of a single dimension.

In addition, it is necessary to discretize the derivative of $f$. For the following discussion let us assume that\footnote{In the following, we employ the Einstein summation convention in case of repeated indices, if appropriate. As common in the literature on dG methods, we extend the orthonormal polynomials $p_{ni}$ by zero outside the $n$-th cell.}
\begin{equation*}
	f(x) = f^{ni}p_{ni}(x).
\end{equation*}
Since we discretize an advection equation we set (see e.g.~\cite{liu2000}) 
\begin{align}
	f_x^{ni} := \hat{f}p_{ni}\vert_{x_{n-1/2}}^{x_{n+1/2}} - \int_{C_n} f \partial_x p_{ni} \,\mathrm{d}x,
    \label{eq:discrete_derivative}
\end{align}
where the numerical flux is denoted by $\hat{f}$ (We will consider an explicit form of the numerical flux later in this section). From this expression, it can be deduced, using integration by parts, that
\begin{equation}
	f_x^{ni} - \int_{C_n} f_xp_{ni}\,\mathrm{d}x = \hat{f}p_{ni}\vert_{x_{n-1/2}}^{x_{n+1/2}} - fp_{ni}\vert_{x_{n-1/2}}^{x_{n+1/2}}. \label{eq:integration-by-parts}
\end{equation}
From the discussion in the previous section we know that we need to employ integration by parts as well as the product rule to show conservation of the invariants under consideration. It is now our goal to investigate if these properties hold true in the present context. For that purpose let us consider the following expression
\begin{equation*}
	\int_{C_n} f_x g \,\mathrm{d}x \approx 
	\int_{C_n} \left( f_x^{ni}p_{ni} \right) \left( g^{nj}  p_{nj} \right) \mathrm{d}x.
\end{equation*}
Using equation \eqref{eq:integration-by-parts} we get
\begin{align*}
	&\int_{C_n} \left( f_x^{ni}p_{ni} \right) \left( g^{nj}  p_{nj} \right) \,\mathrm{d}x \\
	&\qquad=\int_{C_n} f_x g \,\mathrm{d}x + \left( \hat{f}p_{ni}\vert_{x_{n-1/2}}^{x_{n+1/2}} - fp_{ni}\vert_{x_{n-1/2}}^{x_{n+1/2}} \right) \int_{C_n} gp_{ni}\,\mathrm{d}x.
\end{align*}
By employing the orthonormality condition as well as integrating by parts we obtain 
\begin{equation*}
	\int_{C_n} \left( f_x^{ni}p_{ni} \right) \left( g^{nj}  p_{nj} \right) \,\mathrm{d}x
	= \hat{f}g\vert_{x_{n-1/2}}^{x_{n+1/2}} - \int_{C_n} f g_x\,\mathrm{d}x,
\end{equation*}
which upon using equation \eqref{eq:integration-by-parts} gives
\begin{align}
	&\int_{C_n} \left( f_x^{ni}p_{ni} \right) \left( g^{nj}  p_{nj} \right) \,\mathrm{d}x  =
	-\int_{C_n} \left(f^{ni}p_{ni} \right) \left( g_x^{nj}p_{nj} \right)\,\mathrm{d}x \nonumber \\
	&\qquad 
	+ \hat{f}g\vert_{x_{n-1/2}}^{x_{n+1/2}}
	+ \hat{g}f\vert_{x_{n-1/2}}^{x_{n+1/2}}
	- gf\vert_{x_{n-1/2}}^{x_{n+1/2}}. \label{eq:numerical-vs-exact-flux}
\end{align}
This expression relates the error made in the integration by parts to the value at the boundaries and the error in the numerical flux.

In the preceding discussion we have used a generic numerical flux. That is, all the results in this section, up to this point, are true independently of the specific numerical flux under consideration. To be more concrete let us now choose the standard flux, i.e. we assume
\begin{equation}
	\hat{f}(x) = \tfrac{1}{2}\lim_{\epsilon\to 0,\epsilon>0}f(x+\epsilon)+\tfrac{1}{2}\lim_{\epsilon\to 0,\epsilon>0}f(x-\epsilon), \label{eq:standard-flux}
\end{equation}
where for $f\colon [a,b]\to\mathbb{R}$ and periodic boundary conditions, we assume that
\begin{equation}
	\lim_{\epsilon\to 0,\epsilon>0} f(b+\epsilon) = \lim_{\epsilon\to 0,\epsilon>0} f(a+\epsilon), \qquad
	\lim_{\epsilon\to 0,\epsilon>0} f(a-\epsilon) = \lim_{\epsilon\to 0,\epsilon>0} f(b-\epsilon)
\end{equation}
and for homogeneous Dirichlet boundary conditions we assume that
\begin{equation}
	\lim_{\epsilon\to 0,\epsilon>0} f(b+\epsilon) = 0, \qquad
	\lim_{\epsilon\to 0,\epsilon>0} f(a-\epsilon) = 0.
\end{equation}
Let us note that for the standard flux and $P=1$ (i.e.~using a piecewise constant approximation in each cell) the scheme reduces to the classic centered difference scheme of second order. 

Continuing our discussion of the dG method, where we now assume that the numerical flux is given by equation \eqref{eq:standard-flux}, we observe that 
\begin{equation*}
	\sum_n fg\vert_{x_{n-1/2}}^{x_{n+1/2}}
	= \sum_n \left( f_{n+1/2}^{-} g_{n+1/2}^{-} - f^{+}_{n-1/2} g^{+}_{n-1/2} \right),
\end{equation*}
where we have used the notation 
\begin{align*}
	f_{n-1/2}^{+} &= \lim_{\epsilon\to 0, \epsilon>0} f(x_{n-1/2}+\epsilon) \\
	f_{n-1/2}^{-} &= \lim_{\epsilon\to 0, \epsilon>0} f(x_{n-1/2}-\epsilon).
\end{align*}
In addition, the following expression for the numerical flux holds true
\begin{equation*}
	\sum_n \hat{f}g\vert_{x_{n-1/2}}^{x_{n+1/2}} =
	\frac{1}{2} \sum_n 
	\left(
	 	\left[ f_{n+1/2}^{+}+f_{n+1/2}^{-} \right] g_{n+1/2}^{-}
	-	\left[  f_{n-1/2}^{+}+f_{n-1/2}^{-} \right] g_{n-1/2}^{+}
	\right)
\end{equation*}
which, by using summation by parts, can be written as
\begin{equation*}
	\sum_n \hat{f}g\vert_{x_{n-1/2}}^{x_{n+1/2}} =
	\frac{1}{2} \sum_n
		\left[ f_{n-1/2}^{+}+f_{n-1/2}^{-} \right]
		\left[ g_{n-1/2}^{-}-g_{n-1/2}^{+} \right].
\end{equation*}
If we now compute the sum of the three terms in equation \eqref{eq:numerical-vs-exact-flux} together, we immediately see that
\begin{equation*}
	\sum_n 
	\left(
		\hat{f}g\vert_{x_{n-1/2}}^{x_{n+1/2}} +
		f\hat{g}\vert_{x_{n-1/2}}^{x_{n+1/2}}-
		fg\vert_{x_{n-1/2}}^{x_{n+1/2}}
	\right) = 0.
\end{equation*}
This is the desired result which implies that we can use integration by parts for our dG method. Furthermore, we are now in a position to construct a conservative dG scheme by approximating the Poisson bracket as follows
\begin{equation*}
	J = \tfrac{1}{3}(J^{++} + J^{+x} + J^{x+}),
\end{equation*}
where analogous to Arakawa's scheme $J^{++}, J^{+x},$ and $ J^{x+}$ are the dG approximations corresponding to $\{f,g\}=f_xg_y-f_yg_x$, $\{f,g\}=(fg_y)_x-(fg_x)_y$, and $\{f,g\}=(f_xg)_y-(f_yg)_x$, respectively. This scheme then, in accordance with the discussion in the previous section, conserves the vorticity, the kinetic energy, and the enstrophy as the semi-discretized differential is evolved in time.

Based on the discussion in this and the previous sections one might be tempted to conjecture that the constructed scheme is also conservative in the case of homogeneous Dirichlet boundary conditions (as opposed to periodic boundary conditions). 
However, this is not true as, for example, on the right boundary the term $D_x(D_y(f)g)$ does not vanish.
This is confirmed by the numerical experiments conducted in section \ref{sec:simulation}.
In the context of the present discussion let us, however, emphasize that this problem also exists for the finite difference scheme proposed by Arakawa.

\section{Implementation } \label{sec:implementation}
In order to implement and test our theoretical considerations we have to choose a set of orthogonal polynomials and derive the entries of the discrete derivative matrices. 
In this section we consider the Legendre polynomials and show how they are 
employed in our dG scheme. We establish the necessary 
notation and show the close connection to Gauss--Legendre quadrature 
which yields a natural discrete scalar product. Then we derive the 
matrix coefficients of discrete derivatives in one and two dimensions 
for periodic as well as for homogeneous Dirichlet boundary conditions. 

\subsection{ The Legendre polynomials}
First, let us consider the one-dimensional case.
For simplicity and ease of implementation we choose an equidistant grid with $N$ cells of size $h$;
with this choice we are able to construct basis functions of $P(C_n)$, the space of 
polynomials of degree at most $P-1$ on $C_n$, by using orthogonal Legendre polynomials. 
The Legendre polynomials can be recursively defined on $[-1,1]$ by setting
$p_0(x) = 1$, $p_1(x) = x$ and (see e.g.~\cite{AS})
\begin{align}
    (k+1)p_{k+1}(x) = (2k+1)xp_k(x) - kp_{k-1}(x).
    \label{eq:recursion}
\end{align}
The so constructed Legendre polynomials are orthogonal on $[-1,1]$.
We write 
$x^a_j$ and $w_j$, $j=0,\dots,P-1$ denoting the abscissas and weights of 
the Gauss--Legendre quadrature on the interval $[-1,1]$. Then we note that for $k,l=0, \dots, P-1$
\begin{align}
    \int_{-1}^1 p_k(x)p_l(x) \dx = \sum_{j=0}^{P-1} w_jp_k (x^a_j)p_l(x^a_j) = \frac{2}{2k+1}\delta_{kl}, 
    \label{}
\end{align}
 since Gauss--Legendre quadrature is exact for polynomials of degree at most $2P-1$.

The discrete completeness relation can then be written as 
\begin{align}
    \sum_{k=0}^{P-1} \frac{2k+1}{2}w_j p_k(x^a_i)p_k(x^a_j) = \delta_{ij}.
    \label{eq:completeness}
\end{align}
Given a real function $f:[-1,1]\rightarrow \mathbb{R}$ we define $f_j:=f(x^a_j)$ and
\begin{align} \label{eq:ex2}
    \bar f^k := \frac{2k+1}{2}\sum_{j=0}^{P-1}w_j p_k(x^a_j) f_j 
\end{align}
Now let us define the forward transformation matrix by $F^{kj}:=\frac{2k+1}{2}w_jp_k(x^a_j)$ and 
the backward transformation matrix by $B_{kj}:= p_j(x^a_k)$. Then, 
using equation (\ref{eq:ex2}), we get
\begin{subequations}
\begin{align}
    \bar f^k = \sum_{j=0}^{P-1}F^{kj}f_j \\
    f_j = \sum_{k=0}^{P-1} B_{jk}\bar f^k,
\end{align}
\end{subequations}
We call $\bar f^k$ the values of $f$ in $L$-space and $f_j$ the values of $f$ in $X$-space.

Let us now consider an interval $[a,b]$ and an equidistant discretization
by $N$ cells with cell center $x_n$ and grid size $h=\frac{b-a}{N}$; in addition, we set $x_{nj}^a := x_n + \frac{h}{2}x^a_j$.
Given a function $f:[a,b]\rightarrow \mathbb{R}$ we then define
$f_{nj} := f(x^a_{nj})$ and note that
\begin{subequations}
\begin{align}
    \bar{ \vec f} &= (\Eins\otimes F) \vec f \\
    \vec f &= (\Eins\otimes B) \bar{\vec f},
    \label{}
\end{align}
\end{subequations}
where $f_{nj}$ are the elements of $\vec f$,
 $\Eins\in\mathbb{R}^{N\times N}$ is the identity matrix and $F,B\in\mathbb{R}^{P\times P}$. Furthermore, we use
$\otimes$ to denote the Kronecker product which is bilinear and associative. 
The discontinuous Galerkin expansion $f_h$ of a function $f$ in the interval $[a,b]$ can 
then readily be given as
\begin{align}
    f_h(x) = \sum_{n=1}^N \sum_{k=0}^{P-1} \bar f^{nk} p_{nk}(x),
\end{align}
where
\begin{align}
    p_{nk}(x) := \begin{cases}
        p_k\left(  \frac{2}{h}(x-x_n)\right),& \ \text{for } x-x_n\in\left[ -\frac{h}{2}, \frac{h}{2} \right]\\
        0,& \ \text{else}.
    \end{cases}
    \label{}
\end{align}
The use of Legendre polynomials yields a natural approximation of the integrals of $f$
via Gauss--Legendre quadrature
\begin{subequations}
\begin{align}
    \langle f_h,g_h\rangle:=\int_a^b f_hg_h \dx &= \sum_{n=1}^N\sum_{j=0}^{P-1} \frac{hw_j}{2} f_{nj} g_{nj} 
    = \sum_{n=1}^N\sum_{k=0}^{P-1}\frac{h}{2k+1}\bar f^{nk}\bar g^{nk}  \\
    \|f_h\|^2_{L_2} := \int_a^b |f_h|^2 \dx &= \sum_{n=1}^N\sum_{j=0}^{P-1} \frac{h w_j}{2}f_{nj}^2 
    = \sum_{n=1}^N\sum_{k=0}^{P-1} \frac{h}{2k+1}\left(\bar f^{nk}\right)^2. 
    \label{eq:def_norm}
\end{align}
\end{subequations}
With this formulas we have a simple, accurate, and fast 
method to evaluate integrals. This is applied, for example, to compute
errors in the $L_2$-norm.

We now define some useful quantities that simplify our notation (note that $i,j=0,\dots,P-1$) 
\begin{subequations}
    \begin{align}
        S_{ij} &:= \int_{-h/2}^{h/2} p_i\left(\frac{2}{h} x\right)p_j\left(\frac{2}{h} x\right) \dx = \frac{h}{2i+1}\delta_{ij} \\
        T^{ij} &:= S^{-1}_{ij} = \frac{2i+1}{h}\delta_{ij} \\
        W^{ij} &:= \frac{h w_j}{2}\delta_{ij}\\
        V_{ij} &:= W_{ij}^{-1} = \frac{2}{hw_j}\delta_{ij}. 
    \end{align}
    \label{eq:diagonal}
\end{subequations}
Employing these relations we can write
    \begin{align}
		\langle f_h,g_h\rangle =& \vec f^{\mathrm{T}}(\Eins\otimes W)\vec g 
		= \bar{\vec f}^{\mathrm{T}}(\Eins\otimes S)\bar{\vec g} 
    \end{align}
    and \begin{align}
        F = TBW.
        \label{eq:transformation}
    \end{align}
Furthermore, we note that
\begin{subequations}
    \begin{align}
        D_{ij} &:= \int_{-h/2}^{h/2} p_i\left(\frac{2}{h} x\right)\partial_xp_j\left(\frac{2}{h} x\right) \dx\\
		R_{ij} &:= p_i(1)p_j(1) = 1 = R^{\mathrm{T}}_{ij}\\
		L_{ij} &:= p_i(-1)p_j(-1) = (-1)^{i+j} = L^{\mathrm{T}}_{ij}\\
        RL_{ij}&:= p_i(1)p_j(-1) = (-1)^j\\
		LR_{ij}&:= p_i(-1)p_j(1) = (-1)^i = RL^{\mathrm{T}}_{ij}.
    \end{align}
    \label{eq:legendre_operators}
\end{subequations}
In order to compute the elements of $D_{ij}$ we first note that $D_{ij} = 0$ for
$i > j-1$ as $\partial_x p_j(x)$ is a polynomial of degree $j-1$. Then
we use integration by parts to show that 
\begin{align}
	(D+L) = (R-D)^{\mathrm{T}}.
    \label{eq:legendre_derivative}
\end{align}
Therefore, we conclude that $D_{ij} = 1 - (-1)^{i+j}$ for $i\le (j-1)$. 

We introduce the notation (\ref{eq:diagonal}) and (\ref{eq:legendre_operators}) mainly for ease of implementation. If a block-matrix class is written and the
operations $+$, $-$ and $*$ are defined on it, the assembly of the derivative
matrices is simplified to a large extend. 

%

\subsection{ Discretization of the first derivatives}
We are now in a position to write the discretization of $f_x = \partial_x f(x)$, as defined in equation (\ref{eq:discrete_derivative}),
 as a matrix-vector product that can easily be implemented. 
Inserting the standard numerical flux (\ref{eq:standard-flux}) we have\footnote{where once again summation over repeated indices is implied.}
\begin{align}
    \bar f_x^{ni}= T^{ij}&\left[ \quad \frac{1}{2} \left(\bar f^{(n+1)k}p_k(-1)+ \bar f^{nk}p_k(1)\right)p_j(1)\right. \nonumber\\
        &\ \ - \left.\frac{1}{2} \left(\bar f^{nk}p_k(-1) + \bar f^{(n-1)k}p_k(1)\right)p_j(-1) - \bar f^{nk} D_{kj} \vphantom{\frac{1}{2}}\right]
    \label{}
\end{align}
where we used that $p_{nk}(x_{n+1/2})\equiv p_k(1)$ holds true for all $n$.
Together with the previously defined quantities in (\ref{eq:legendre_operators}) we can write
\begin{align}
	\bar{\mathbf f}_x  &= (\Eins\otimes T)\circ \left[ \frac{1}{2}(\Eins^+\otimes RL + \Eins \otimes (D-D^{\mathrm{T}}) - \Eins^-\otimes LR)\right] \bar{\mathbf f}\nonumber \\
    &=: (\Eins\otimes T) \circ \bar M_x^{per} \bar {\mathbf f},
    \label{eq:discrete_der}
\end{align}
using $D+D^{\mathrm{T}} = R-L$ from equation (\ref{eq:legendre_derivative})
and 
\begin{align}
    \Eins^{-}f^{n} &:= f^{n-1}\\
    \Eins^{+}f^{n} &:= f^{n+1}.
    \label{eq:operator_one}
\end{align}
We define
\begin{align*}
    M_x^{per} := (\Eins\otimes F^T)\circ \bar M_x^{per}\circ (\Eins\otimes F).
    \label{}
\end{align*}
If our coefficients are given in $X$-space we note with the help of equation (\ref{eq:transformation})
\begin{align}
	\mathbf f_x 
 = (\Eins\otimes V)\circ M_x^{per} \mathbf f,
    \label{eq:matrix_xspace}
\end{align}
where $\bar M_x^{per}$ and $M_x^{per}$ are skew-symmetric matrices.

From equation (\ref{eq:discrete_der}) we are now able to show the matrix representation of the one-dimensional discrete derivative for 
periodic and homogeneous Dirichlet boundary conditions that can be used
in the implementation
\begin{align}
    \bar M_x^{per} = \frac{1}{2}\begin{pmatrix}
		(D-D^{\mathrm{T}}) & RL      &    &   & -LR \\
		-LR  & (D-D^{\mathrm{T}}) & RL &   &     \\
             &  -LR    & \dots   &   &     \\
             &         &    & \dots  & RL    \\
			 RL &         &    & -LR&(D-D^{\mathrm{T}}) 
    \end{pmatrix}
    \label{}
\end{align}

\begin{align}
    \bar M_x^{dir} = \frac{1}{2}\begin{pmatrix}
		(D-D^{\mathrm{T}}) & RL      &    &   & 0 \\
		-LR  & (D-D^{\mathrm{T}}) & RL &   &     \\
             &   -LR   & \dots   &   &     \\
             &         &    & \dots  & RL    \\
			 0  &         &    & -LR&(D-D^{\mathrm{T}})
    \end{pmatrix}.
    \label{}
\end{align}
Note that for $P=1$ we recover the centered finite difference approximation of the first derivative. 

In this notation the local character of the dG method is apparent.
To compute the derivative in one cell we only use values of neighboring
cells. Therefore, the method is well suited for parallelization which we
will exploit in our implementation.
The generalization to two dimensions is straightforward. If we 
operate on the product space $[x_0, x_1]\times[y_0,y_1]$, all the 
matrices derived above can be readily extended via the appropriate Kronecker products.
The space complexity of 
the matrices derived is $\mathcal{O}(P^2 N)$ in one and $\mathcal{O}(P^3N^2)$ in two dimensions.

Finally, let us remark that we found it practical to always operate on 
coefficients in $X$-space, i.e.~we use equation (\ref{eq:matrix_xspace}) 
for our implementations and thus use $\vec f$ rather than $\bar{\vec f}$ to represent the approximation. The function products, as necessary in the scheme proposed, are easily computed coefficient-wise in $X$-space, i.e.~we use
\begin{align}
    (fg)_{ni}=f_{ni}g_{ni}
    \label{}
\end{align}
to represent the corresponding products.

%
%

\section{Numerical experiments I } \label{sec:simulation}
In this section we present numerical experiments that verify our 
theoretical considerations. In addition, we briefly describe the implementation and efficiency of our GPU code and numerically 
show that the invariants of the Jacobian are indeed conserved for
periodic boundary conditions.
\subsection{Implementation}
Our implementation
follows modern C++ design principles \cite{Furnish1998}, 
which enable the separation of numerics and optimization through 
template metaprogramming. 
Note that only BLAS level 1 and 2 function kernels need
to be written in order to implement our dG code. 
At this time
both an OpenMP parallelized CPU and a GPU backend has been implemented based on the use of 
CUDA thrust and the cusp library \cite{Thrust, Cusp}. 
The advantage of this approach is that we can switch backends and support additional architectures without 
having to change the high level implementation of the numerical method. 

Although there is still room for optimization, both on the GPU as 
well as on the CPU side, we observe good parallel efficiency 
of the code due to the 
 high degrees of parallelism intrinsic to the discontinuous Galerkin discretization.
 We compare the average run time of a single evaluation of the Arakawa bracket 
 on a Nvidia GeForce GTX 570 to an Intel(R) Xeon(R) CPU E3 1225 V2 using 4 threads. 
 For the computation under consideration we fix $N=100$ and vary $P$. The speedup is given by $S=T^{CPU}/T^{GPU}$ and 
 the result is shown in Table \ref{tab:performance}. We observe a 
 speedup of $2$ to $6$ in favor of the GPU dependent on problem size.
 As usual in scientific applications we employ double precision floating point numbers both on the CPU as well as on the GPU.

\begin{table}[htbp]
\begin{center}
\begin{tabular}{|c|c|c|c|c|}
\hline
P & $T^{GPU}/ms$ & $T^{CPU}/ms$ & speedup \\ \hline
1 & 0.05 & 0.08 & 1.5 \\ \hline
2 & 0.26 & 0.52 & 2.0 \\ \hline
3 & 0.71 & 3.85 & 5.4 \\ \hline
4 & 1.47 & 9.08 & 6.2 \\ \hline
\end{tabular}
\end{center}
\caption{Average run time on NVIDIA's GeForce GTX 570 compared to Intel's Xeon CPU E3 1225 V2. Problem size is $\mathcal {O}(P^3 N^2)$, $N$ is fixed to $100$.  }
\label{tab:performance}
\end{table}

\subsection{Conservation properties}
Let us first consider the functions
\begin{subequations}
\begin{align}
    f(x,y) &= \sin(x)\cos(y) \\
	g(x,y) &= \mathrm{e}^{0.1(x+y)} 
\end{align}
\end{subequations}
on the domain $[0, \pi] \times [0, \pi]$.
We fix $P=3$ and $N_x = N_y = 112$.
Then we numerically compute the Jacobian and   
compare the conservation
properties for $J^{++}$, $J^{x+}$, $J^{+x}$ as defined in equation (\ref{eq:J}) using periodic boundary conditions. In addition, we perform a computation with the
discretized Jacobian that has been developed in this paper (for both periodic and homogeneous Dirichlet boundary conditions). 
The result is shown in Table \ref{tab:arakawa_conservation}. 
\begin{table}[htbp]
\begin{center}
\begin{tabular}{|c|c|c|c|}
\hline
 & $\int J\dA$ & $\int f J \dA$ & $\int g J\dA$ \\ \hline
$J^{++}$ & 1.67E-016 & 0.068 & -0.038 \\ \hline
$J^{+x}$ & -3.89E-016 & -0.068 & -6.66E-016 \\ \hline
$J^{x+}$ & 5.55E-016 & 4.16E-016 & 0.038 \\ \hline
periodic & 3.89E-016 & -6.38E-016 & 7.77E-016 \\ \hline
Dirichlet & -0.19 & 0.034 & -0.79 \\ \hline
\end{tabular}
\end{center}
\caption{Conservation properties of various discretization methods as defined in equation (\ref{eq:J}).}
\label{tab:arakawa_conservation}
\end{table}

We observe that only the proposed discretization conserves all of the invariants in the case of periodic boundary conditions. In addition, we see that for all three invariants this conservation is violated for homogeneous Dirichlet boundary conditions (as was expected from the theoretical considerations in section \ref{sec:dG}).

\section{ Numerical experiments II} \label{sec:simulation2}
In this section we present numerical simulations of
the two-dimensional incompressible 
Navier--Stokes equations (\ref{eq:vorticity}).
For the discretization of the Laplacian we use the local discontinuous 
Galerkin method developed by \cite{Cockburn1998}, which was later 
shown to have superconvergence properties for elliptic problems
on Cartesian grids \cite{Cockburn2001}.
This approach is combined with the dG scheme proposed in this paper. The result is an implementation that
completely relies on a dG discretization in space, i.e.~we do not employ any 
continuous finite element methods to solve the Poisson equation (as is done in \cite{liu2000}, for example). 
In time we use a standard explicit three-step Adams--Bashforth method. 
The dG discretization of the Poisson equation leads to a symmetric matrix equation that
we solve by using a conjugate gradient method. An initial guess is computed by extrapolating the solutions of the last two time steps. 
Since the conjugate gradient method is based on matrix-vector 
multiplications and vector additions, we can readily employ our previously 
implemented routines without modification. 

We will first numerically evaluate the order 
of our method and then analyze the conservation properties of our
numerical scheme in the context of the so called Lamb dipole. 

\subsection{Order of the dG method}
Let us note that equations (\ref{eq:vorticity}) have an analytical solution for the following initial condition
\begin{align}
	\omega( x, y, 0 ) = 2\sin(x)\sin(y)
\end{align}
on the domain $[0,2\pi]\times[0,2\pi]$. This solution is given by 
\begin{align}
	\omega( x, y, t) = 2\sin(x)\sin(y)e^{-2Dt} = 2\psi(x,y,t)
	\label{}
\end{align}
which fulfills periodic as well as Dirichlet boundary conditions.
We use this to numerically test the order of our discretization. 
We integrate from $0$ to $2$ respecting the CFL condition and compute 
absolute errors in the $L_2$-norm, i.e.
\begin{align}
    \eps = \| \vec \omega - \vec \omega_h\|_{L_2},
\label{}
\end{align}
for various values of $P$ and $N$ using equation (\ref{eq:def_norm}).
The results are shown in Table \ref{tab:accuracy}. We observe 
convergence of order $P$ in all cases with $P\geq 2$. 
For $P=1$ and periodic boundaries we fall back to the finite difference
discretization which explains the order $2$ convergence in the finite
diffusion case. For $D=0$ Poisson's equation happens to be exactly solved which leads to 
an exact cancellation of errors. 
This, however, is not the case for a different numbers of cells ($15^2$ for example).
In the case of Dirichlet
boundary conditions we only achieve first order convergence. 

\begin{table}[htbp]
\begin{center}
\Rotatebox{90}{%
\begin{tabular}{|c|c|c|c|c|c|c|c|c|}
\hline
 & \multicolumn{4}{|c|}{$D=0$} & \multicolumn{4}{|c|}{$D=0.01$} \\ \hline
 & \multicolumn{2}{|c|}{Periodic BC} &\multicolumn{2}{|c|}{Dirichlet BC}  & \multicolumn{2}{|c|}{Periodic BC} &\multicolumn{2}{|c|}{Dirichlet BC} \\ \hline
 \# of cells & $L^2$ error & order & $L^2$ error & order & $L^2$ error & order & $L^2$ error & order \\ \hline
 \multicolumn{9}{|c|}{ $P=1$ }  \\ \hline
$16^2$ & 0.00E+00 & - & 4.00E-01 & - & 2.63E-03 & - & 3.79E-01 & - \\ \hline
$32^2$ & 0.00E+00 & - & 3.30E-01 & 0.28 & 6.59E-04 & 2.00 & 3.07E-01 & 0.30 \\ \hline
$64^2$ & 0.00E+00 & - & 2.00E-01 & 0.72 & 1.79E-04 & 1.88 & 1.90E-01 & 0.69 \\ \hline
$128^2$ & 0.00E+00 & - & 1.10E-01 & 0.86 & 4.48E-05 & 2.00 & 1.00E-01 & 0.93 \\ \hline
 \multicolumn{9}{|c|}{ $P=2$ }  \\ \hline
$16^2$ & 5.96E-02 & - & 5.46E-02 & - & 5.63E-02 & - & 5.71E-02 & - \\ \hline
$32^2$ & 1.48E-02 & 2.01 & 1.38E-02 & 1.98 & 1.49E-02 & 1.92 & 1.51E-02 & 1.92 \\ \hline
$64^2$ & 3.74E-03 & 1.98 & 3.60E-03 & 1.94 & 3.83E-03 & 1.96 & 3.87E-03 & 1.96 \\ \hline
$128^2$ & 9.44E-04 & 1.99 & 9.20E-04 & 1.97 & 9.74E-04 & 1.98 & 9.80E-04 & 1.98 \\ \hline
 \multicolumn{9}{|c|}{ $P=3$ }  \\ \hline
$16^2$ & 2.59E-03 & - & 2.47E-03 & - & 1.80E-03 & - & 1.69E-03 & - \\ \hline
$32^2$ & 3.36E-04 & 2.95 & 3.27E-04 & 2.92 & 2.32E-04 & 2.96 & 2.25E-04 & 2.91 \\ \hline
$64^2$ & 4.34E-05 & 2.95 & 4.28E-05 & 2.93 & 2.95E-05 & 2.98 & 2.91E-05 & 2.95 \\ \hline
$128^2$ & 5.51E-06 & 2.98 & 5.47E-06 & 2.97 & 3.73E-06 & 2.98 & 3.70E-06 & 2.98 \\ \hline
 \multicolumn{9}{|c|}{ $P=4$ }  \\ \hline
$16^2$ & 3.16E-05 & - & 3.33E-05 & - & 4.33E-05 & - & 4.49E-05 & - \\ \hline
$32^2$ & 1.84E-06 & 4.10 & 1.91E-06 & 4.12 & 2.78E-06 & 3.96 & 2.84E-06 & 3.98 \\ \hline
$64^2$ & 1.11E-07 & 4.05 & 1.13E-07 & 4.08 & 1.76E-07 & 3.98 & 1.78E-07 & 4.00 \\ \hline
$128^2$ & 6.79E-09 & 4.03 & 6.84E-09 & 4.05 & 1.11E-08 & 3.99 & 1.12E-08 & 3.99 \\ \hline
\end{tabular}
}%
\end{center}
\caption{Accuracy for the dG method proposed for various number of cells $N$ and polynomial degrees in each cell $P$.}
\label{tab:accuracy}
\end{table}

\subsection{ The Lamb dipole}
In order to test the conservation properties of the scheme developed in this paper
we integrate the Lamb dipole \cite{Nielsen1997}.
The Lamb dipole is a stationary solution to the Euler equations (equation (\ref{eq:vorticity}) with $D=0$) in an infinite domain:
\begin{align}
	\omega = \begin{cases}
		\frac{2\lambda U}{J_0(\lambda R)} J_1(\lambda r)\cos \theta,& \ \ r<R,\\
		0,& \ \ r>R,
	\end{cases}
	\label{eq:lamb}
\end{align}
where $r$ and $\theta$ are given in polar coordinates.
The parameter $U$ denotes the velocity of the dipole, $J_i$ is 
the $i$-th Bessel function of the first kind, $R$ is the radius of the dipole,
and $\lambda R$ is the first zero of $J_1$ which is given by
\begin{align}
    \lambda R = 3.83170597020751231561.
    \label{}
\end{align}
The dipole 
will be at rest in the frame of reference moving with constant velocity $U$ in the negative $y$ direction.
A finite box will slightly decrease the actual velocity of the dipole.

For the Lamb dipole the total vorticity exactly vanishes, while energy and enstrophy are 
exactly conserved. 
The integrals can be naturally evaluated using Gauss--Legendre quadrature in our numerical scheme.
We consider the following measure of error in these variables
\begin{subequations}
	\begin{align}
		\eps_V &:= V \\
		\eps_\Omega &:= \left|\frac{\Omega-\Omega_0}{\Omega_0}\right| \\
		\eps_E &:= \left|\frac{E-E_0}{E_0}\right|.
		\label{}
	\end{align}
	\label{}
\end{subequations}
Let us fix $l_x = l_y = 1$, $U=1$ and $R = 0.1$.
In order to evaluate the conservation properties of our proposed
scheme in connection with a time-stepping scheme,
we choose $P=1$ and $N=200$ and integrate from $0$ to $0.01$ with 
the $K$-step Adams--Bashforth formula.
The result is shown in Table \ref{tab:timesteps}.
We indeed recover order $K$ for the $K$-step algorithm, except for $K=2$, where error cancellations might explain the $3$rd order convergence.
\begin{table}[htbp]
\begin{center}
\begin{tabular}{|c|c|c|c|c|}
\hline
\# of time steps & $\eps_\Omega$ & order & $\eps_E$ & order \\ \hline
\multicolumn{5}{|c|}{ $K=1$ (Euler method) }  \\ \hline
10 & 3.30E-003 & - & 1.73E-003 & - \\ \hline
20 & 1.65E-003 & 1.00 & 8.62E-004 & 1.00 \\ \hline
40 & 8.22E-004 & 1.00 & 4.31E-004 & 1.00 \\ \hline
80 & 4.11E-004 & 1.00 & 2.15E-004 & 1.00 \\ \hline
\multicolumn{5}{|c|}{ $K=2$ }  \\ \hline
10 & 2.21E-006 & - & 5.49E-007 & - \\ \hline
20 & 2.71E-007 & 3.03 & 6.65E-008 & 3.04 \\ \hline
40 & 3.33E-008 & 3.03 & 7.82E-009 & 3.09 \\ \hline
80 & 4.03E-009 & 3.05 & 8.55E-010 & 3.19 \\ \hline
\multicolumn{5}{|c|}{ $K=3$ }  \\ \hline
10 & 2.76E-006 & - & 7.12E-007 & - \\ \hline
20 & 3.78E-007 & 2.87 & 9.73E-008 & 2.87 \\ \hline
40 & 4.93E-008 & 2.94 & 1.27E-008 & 2.94 \\ \hline
80 & 6.29E-009 & 2.97 & 1.62E-009 & 2.97 \\ \hline
\multicolumn{5}{|c|}{ $K=4$ }  \\ \hline
10 & 6.67E-010 & - & 4.96E-009 & - \\ \hline
20 & 3.22E-010 & 1.05 & 3.45E-010 & 3.85 \\ \hline
40 & 3.95E-011 & 3.02 & 2.31E-011 & 3.90 \\ \hline
80 & 3.19E-012 & 3.63 & 1.49E-012 & 3.95 \\ \hline
\end{tabular}
\end{center}
\caption{Error in the enstrophy and kinetic energy for an Adams-Bashforth multistep method of order $K$. The Lamb dipole is integrated to a final time of $0.01$. The table shows that the error in the conserved quantities is only due to the error in the time stepping scheme.}
\label{tab:timesteps}
\end{table}

We conclude that the errors originate in the time stepping algorithm only. Let us further remark that
for higher order dG methods the CFL condition usually 
restricts the time step in such a manner that the errors in the conserved quantities are comparable to machine precision. 

\begin{figure}[htpb]
    \begin{center}
        \subfloat[ $J^{++}$]{
            \label{}
            \includegraphics[width= 0.4\textwidth]{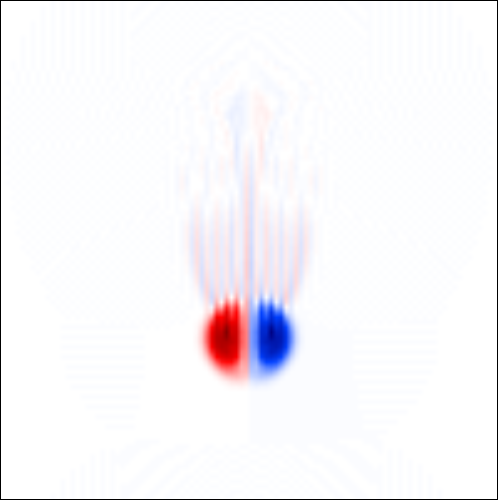}
        }
		\subfloat[ $\tfrac{1}{3}(J^{++}+J^{+x}+J^{x+})$ ]{
            \label{}
            \includegraphics[width= 0.4\textwidth]{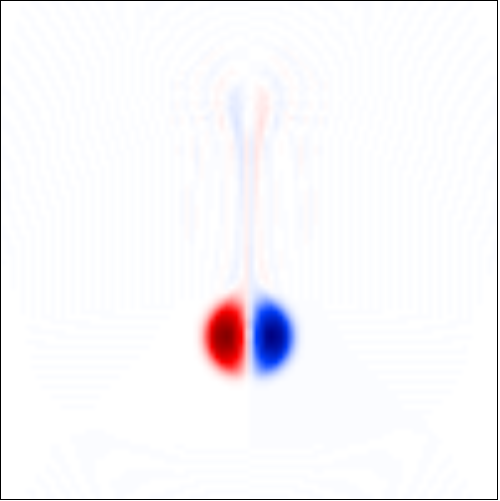}
        }
    \end{center}
    \caption{ Vorticity plot of the Lamb Dipole at $t=0.5$ for $P=1$ and $N=100$. Both schemes show visible errors, but the discretization developed in this paper retains the dipolar form much better
    than the naive discretization.}
    \label{fig:low_order}
\end{figure}

\begin{figure}[htpb]
    \begin{center}
        \subfloat[ $J^{++}$]{
            \label{}
            \includegraphics[width= 0.4\textwidth]{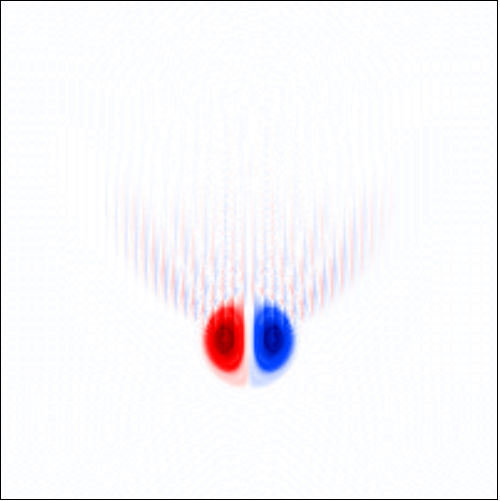}
        }
        \subfloat[ $\tfrac{1}{3}(J^{++}+J^{+x}+J^{x+})$ ]{
            \label{}
            \includegraphics[width= 0.4\textwidth]{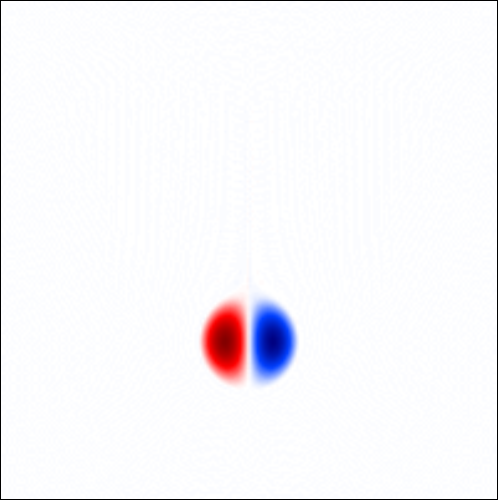}
        }
    \end{center}
    \caption{ Vorticity plot of the Lamb Dipole at $t=0.5$ for $P=4$ and $N=50$. Also in this high order simulation our novel discretization retains the dipolar form much better
    than the naive discretization.}
    \label{fig:high_order}
\end{figure}

Finally, we present plots of the Lamb dipole at $t=0.5$ for low and high 
order discretizations in Figure \ref{fig:low_order} and Figure \ref{fig:high_order}, respectively.
We compare the simulation of the scheme proposed in this paper to
the straightforward discretization of $J^{++}$ as given by equation (\ref{eq:J}).
The Lamb dipole should exactly keep its form during a simulation. We observe that
for both the low and high order discretization the scheme developed in this paper achieves superior results in respecting the 
structure of the dipole. For the naive scheme unphysical oscillations are 
observed.


\section{Conclusion } \label{sec:conclusion}

We have developed a novel conservative space discretization of the two-dimensional
Poisson bracket combining the well-known Arakawa scheme with a high
order discontinuous Galerkin method. Together with an existing dG
discretization for the Laplacian we were able to discretize the 
two dimensional incompressible Navier--Stokes and Euler equations. Simulations
confirm the high order and the conservative properties of our method. In addition, we have demonstrated that an efficient parallel implementation on both CPUs as well as GPUs can be attained.

\section*{Acknowledgements}

The authors are grateful to Prof. A. Ostermann of the Department of Mathematics 
and to Prof. A. Kendl of the Institute for Ion Physics and Applied Physics, 
University of Innsbruck, for valuable discussions and comments. 
LE was supported by the Austrian Science Fund (FWF) project P25346.
MW was partly supported by the Austrian Science Fund (FWF) project W1227-N16, and
by the European Commission under the Contract of Association between EURATOM
and \"OAW, carried out within the framework of the European Fusion Development
Agreement (EFDA). The views and opinions expressed herein do not necessarily
reflect those of the European Commission.

\bibliography{paper}
\bibliographystyle{model1-num-names}

\end{document}